\documentclass[12pt, a4paper]{article}

\usepackage[version=3]{mhchem} 
\usepackage[T1]{fontenc}       
\usepackage{verbatim}
\usepackage{subcaption}
\usepackage{amsmath}
\usepackage[numbers,sort]{natbib}
\usepackage{graphicx}
\usepackage{authblk}
\usepackage[english]{babel}

\title
  {Line tension and wettability of nanodrops on curved surfaces}
\author[1]{Shantanu Maheshwari}

\author[1]{Martin van der Hoef}

\author[1,2]{Detlef Lohse\thanks{d.lohse@utwente.nl}}

\affil[1]{Physics of Fluids, Department of Science and Technology, Mesa+ Institute, and J. M. Burgers Centre for Fluid Dynamics, University of Twente, P.O. Box 217, 7500 AE, Enschede, The Netherlands}
\affil[2]{Max Planck Institute for Dynamics and Self-Organization, 37077, Göttingen, Germany}
\date{}

\begin{document}
\maketitle

\begin{abstract}

In this paper we study the formation of nanodrops on curved surfaces (both convex and concave) by means of molecular 
dynamics simulations, where the particles interact via a Lennard-Jones potential. We find that the contact angle is not affected
by the curvature of the substrate, in agreement with previous experimental findings. This means that the change in curvature of the drop
in response to the change in curvature of the substrate can be predicted from simple geometrical considerations, under the
assumption that the drop's shape is a spherical cap, and that the volume remains unchanged through the curvature. The resulting prediction is in 
perfect agreement with the simulation results, for both convex and concave substrates. In addition, we calculate the line tension, namely
by fitting the contact angle for different size drops to the modified Young equation. We find that the line tension for concave 
surfaces is larger than for convex surfaces, while for zero curvature it has a clear maximum. This feature is found to be correlated with
the number of particles in the first layer of the liquid on the surface.
\end{abstract}

\section{Introduction}

The line tension is a key property for understanding the behavior of nanodrops, and thereby of great
technological relevance for lithography techniques, or in micro and nanofluidics\cite{pethica1977,amirfazli2004,peters2000,lopes2001}.
On {\em{curved}} surfaces the line tension is of great significance
for froth floatation, microporous solid and condensation on nanorods\cite{he2007}.
Due to the small magnitude of line tension, it can affect wetting properties at the nanoscale without having any effect on micro or macroscale.
Understanding, and hence predicting the line tension of nanodrops is non-trivial.
Although there have been a number of theoretical and experimental studies on the subject, 
up to recently there was no consensus even on the sign nor on the order of magnitude\cite{schimmele2007}.

\begin{figure}
  \includegraphics[width=150mm]{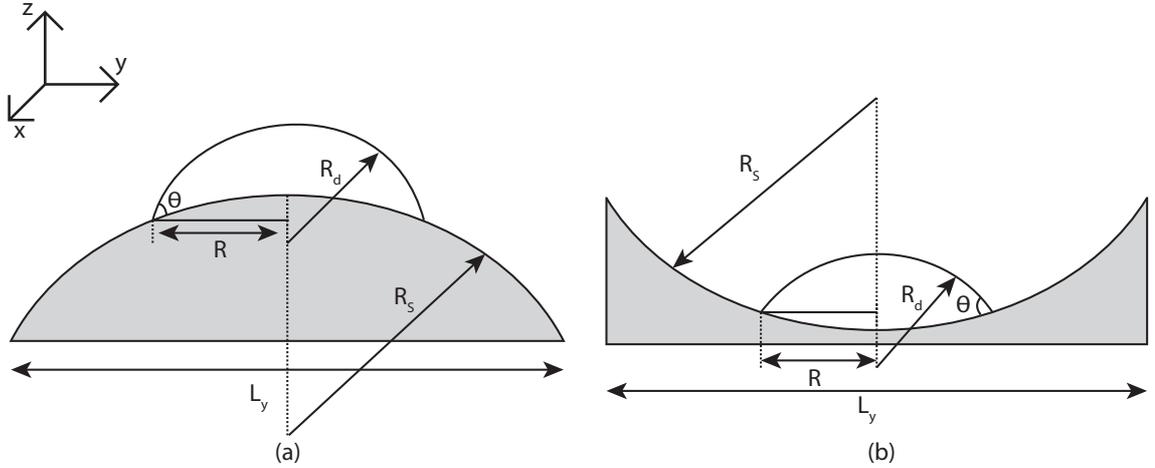}
  \caption{Definition of the various geometrical parameters used in this work, for (a) the convex surface, and (b) the concave surface.
   Three different radi of curvatures can be identified: radius of surface curvature $\rm{R_{S}}$,
 radius of drop curvature $R_{d}$ and the radius of curvature of the contact line $R$. Note that we define $\rm{R_{S}}$ such that for the convex surface it has a positive value, and for the concave surface a negative value.}
  \label{fig1}
\end{figure}

The concept of line tension was introduced more than a century ago by \citeauthor{gibbs}\cite{gibbs}, who concluded that interactions at the three 
phase contact line cannot be explained by surface free energies of each pair of phases alone. He defined the line tension as the excess 
free energy per unit length of a contact line of three phases, analogous to surface tension, which is the excess free energy per unit area. 
In \citeyear{harkins1937}, \citeauthor{harkins1937}\cite{harkins1937} 
managed to theoretically calculate the order of magnitude of line tension from 
the relation between latent heat of vaporisation and the free, latent and total energy of the three phase contact line.
In \citeyear{pethica1977}, \citeauthor{pethica1977}\cite{pethica1977} defined line tension 
for a liquid drop on an ideal solid surface. He included the
line tension in the conditions for equilibrium in the free energy expression, which, when minimised with respect to the contact angle at constant volume,
gives the so-called modified Young equation:

\begin{equation}
  \cos\theta = \frac{\gamma_{SV} - \gamma_{SL}}{\gamma_{LV}} - \frac{\tau/\gamma_{LV}}{R}  = \cos\theta_{Y} - \frac{\tau/\gamma_{LV}}{R},\label{myl}
\end{equation}
\nolinebreak where $\theta$ is the contact angle and $R$ is the radius of curvature of the
contact circle of the liquid drop on an ideal (chemically and geometrically homogenous) solid surface when it is in equilibrium with its own vapour, 
as illustrated in Figure \ref{fig1}. In eq \ref{myl}, $\tau$ is 
the line tension, $\theta_{Y}$ Young contact angle and
$\gamma_{SL}, \gamma_{SV}, \gamma_{LV}$ are the solid-liquid, solid-vapour, and liquid-vapour surface tension, respectively.
In eq \ref{myl}, we have not considered the effect of curvature of the liquid-vapour interface on the surface tension\cite{tolman1949}, because if
these effects become comparable in magnitude to the line tension then the measured $\tau$ cannot be considered as 'pure' line tension, but as an apparent
line tension\cite{schimmele2007,schimmele2009}. 

The magnitude of the line tension has been calculated
from the free energy associated with the three phase contact line using density functional theory\cite{getta1998,bauer1999}
or a model based on the interface displacement\cite{dobbs1993,churaev1982}. Most of these theoretical analyses predict a value 
of the line tension in the range of $10^{-12}$ N to $10^{-11}$ N. Experimental 
investigations show that the order of magnitude of an effective line tension varies from $10^{-5}$ N to $10^{-12}$ N with both positive and negative sign
\cite{gaydos1987,li1990,amirfazli1998,heim2013,marmur1997,schimmele2007,amirfazli2004}, , which basically only shows
how challenging it is to measure the line tension experimentally. The 
primary reason for this variation is the contact angle hysteresis caused by surface heterogeneities, either geometric or chemical,
which are always present for an actual experimental situation. But also the extremely low size range of the drops makes it very difficult to 
measure the line tension experimentally. For example, in case of water drops with a surface tension of $0.072$ N/m and 
a line tension of $10^{-11}$ N (the most consistent 
order of magnitude reported in literature), the line tension becomes significant only for contact line curvatures of around 5 nm\cite{lohse2015rmp}.

Studies have not been confined to flat surfaces.
\citeauthor{extrand2008}\cite{extrand2008} have investigated the dependence of contact angle on surface curvatures experimentally, 
however for drop sizes in the micro- and millimeter range. \citeauthor{marmur2002}\cite{marmur2002} have calculated the  
line tension on spherical surfaces from theoretical calculations but it lacks validation from experimental or simulation data.

Apart from theoretical and experimental research, there have also been a number of studies on the line tension 
and measurement of contact angle using molecular dynamics 
simulations\cite{weijs2011pof,shi2009,ingebrigtsen2007,halverson2010}.
Such simulations have the advantage that the line tension can be calculated with relatively large accuracy, and also the heterogeneities can be well controlled.
\citeauthor{shi2009}\cite{shi2009} studied the behaviour of the contact angle of drops on a plane solid substrate as a function of
temperature and force field parameters, using a simple Lennard-Jones model (as described in next section), as well as a more advanced model potential for water. 
\citeauthor{ingebrigtsen2007}\cite{ingebrigtsen2007} have calculated the contact angles for drops of
different sizes and observed that the contact angle from MD simulations disagree with Young contact angle for nanodrops which have a very small contact angle. 
\citeauthor{weijs2011pof}\cite{weijs2011pof} have analysed the effect of line tension by measuring contact angles of droplets on a plane substrate by 
varying the droplet size, and found that the line tension decreases with increasing $\theta_Y$.

In the present paper, we have extended the simulation by \citeauthor{weijs2011pof}\cite{weijs2011pof} to curved substrates.
Simulations have been performed in 3D, that is, for spherical drops, and in quasi 2D (where one dimension is considerably smaller than the other two), 
to which we refer as cylindrical 
drops. Cylindrical drops give the value of Young contact angle ($\theta_{Y}$), as the contact line is free from any curvature and hence the contact angle does not 
change with the size of the drop.
The effect of the curvature on the contact line can then be predicted by calculating the contact angle for spherical drops of different sizes and comparing it with Young
contact angle ($\theta_{Y}$). In this way,
we have systematically studied the effect of surface curvature on magnitude of line tension and wettability of nanodrops on curved surfaces.

\section{Numerical method}
\subsection{Molecular dynamics simulations of nanodrops}

Molecular Dynamics (MD) simulations were performed to simulate the drop on a solid substrate 
for which we used the open source code GROMACS\cite{gromacs}. Two kind of particles were used in the 
simulations; solid substrate particles which are held fixed in a fcc lattice setting during whole simulation, and 
liquid/vapour particles, which are free to move, and in the equilibrium state form in a liquid drop 
on the solid substrate, and a vapour phase filling the remaining volume. The interaction between the particles is described by Lennard-Jones potential:
\begin{equation}
\phi^{\rm{LJ}}_{ij} (r) = 4\epsilon_{ij}\Big[\Big({\frac{\sigma_{ij}}{r}}\Big)^{12} - \Big({\frac{\sigma_{ij}}{r}}\Big)^6\Big], \label{lj}
\end{equation} 
\nolinebreak in which $\epsilon_{ij}$ is the interaction strength between particles $i$ and $j$, and $\sigma_{ij}$ is the characteristic 
size of particles, which is set to a value $\sigma=0.34$ nm for all interactions.
The potential is truncated at a relatively large cut-off radius of $r_{c} = 5\sigma$. The time step for updating the particle 
velocities and positions was set at  $dt = \sigma\sqrt(m/\epsilon_{LL})/400$, where $m$ is mass of the 
particles and $\epsilon_{LL} = 3$ kJ/mol is the Lennard-Jones interaction parameter for the liquid phase.
Simulations have been performed in an $NVT$ ensemble where the temperature is fixed at 
$300 K$, which is below the critical point for the Lennard-Jones parameters ($\sigma$, $\epsilon_{LL}$) that we have used. 
Periodic boundary conditions have been employed in all three directions.
We have studied two different kinds of 
systems to examine the effect of line tension: quasi-2D and 3D. In quasi-2D, the system size in one dimension is substantially
less than the size of system in other two dimensions. The typical dimension of the system is 
$10.5\sigma$ in the $x$-direction and around $150\sigma$ in the
$y$ and $z$-direction, where the $x,y,z$ directions are defined in Figure \ref{fig1}. In 3D, the system size in all three 
directions is of the same order of magnitude. The system size is such that in all 
cases the distance between a drop and its neighbouring image is at least $80\sigma$. In all simulations, the overall number density is kept constant. 
The equilibrium contact angle (or wettability) of the liquid drop on the solid substrate was varied by changing the interaction 
strength between solid and liquid particles ($\epsilon_{SL}$), from $1.0$ to $2.0$ kJ/mol. In all simulations 
the liquid drop was found to be in equilibrium with its own vapour, where the liquid and vapour density was found to be in
close agreement with the theoretical result as obtained from the equation of 
state of the Lennard-Jones fluid\cite{johnson1993}.

In the simulations, the line tension is evaluated in terms of the tension length ($\ell$) defined as $\ell=-\tau/\gamma_{LV}$, 
which is of the order of magnitude of the molecular scale. We have calculated the magnitude of the line tension length
along the lines of \citeauthor{weijs2011pof}\cite{weijs2011pof}, by measuring the 
equilibrium $\theta$ for different size drops, and fitting a straight line to $\cos\theta$ vs. $1/R$, where the slope is then equal to $\ell$. 
Repeating these calculations for various values of the LJ parameters then gives the tension length ($-\tau/\gamma_{LV}$) as a function of $\theta_{Y}$.

\begin{figure}
\includegraphics[width=150mm]{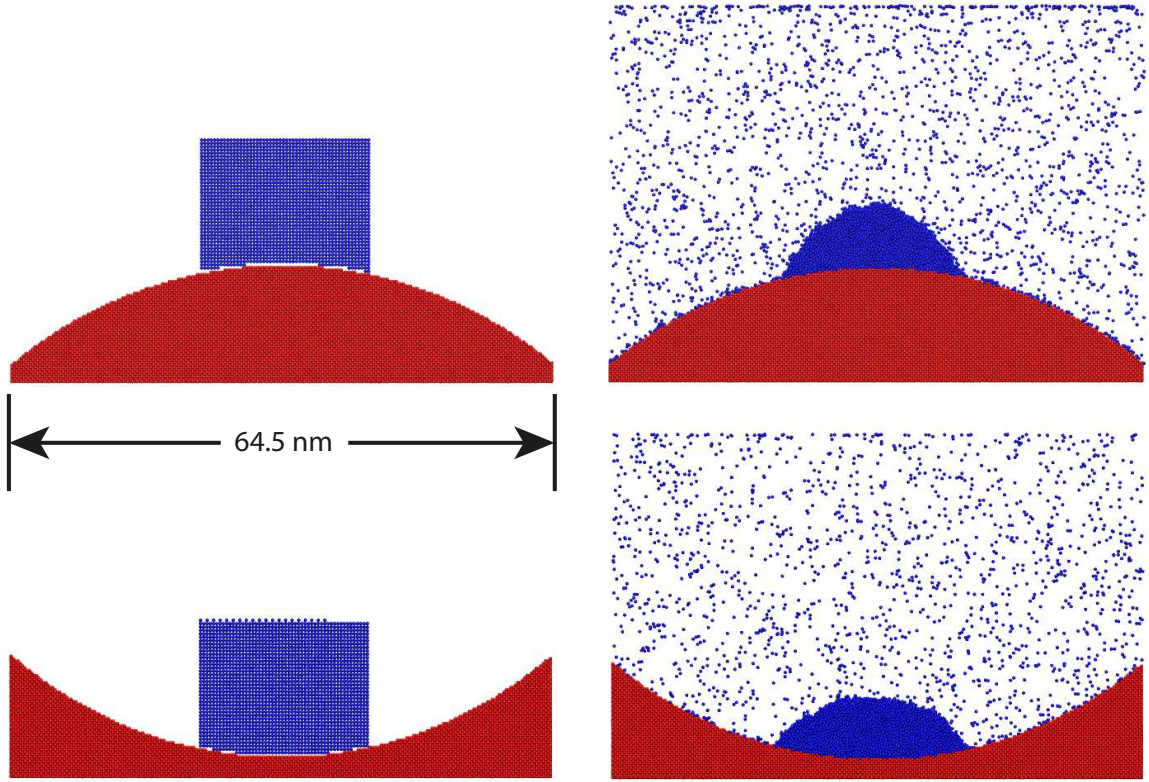}
\caption{Example of the initial configuration (left) and the final steady-state configuration (right) after $5\times10^7$ time steps of the nanodrop simulations for convex (top) and concave (bottom) surfaces. }
\label{fig2}
\end{figure}

Two kinds of surface curvatures were used in this study as shown in Figure \ref{fig2}: a curved outward (or convex) surface defined as positive curvature and 
a curved inward (or concave) surface defined as negative curvature. 
In order to keep the overall number density constant, we have scaled the system 
dimensions while increasing the number of particles in the simulations. We have also scaled the radius of 
curvature of the surface according to the system dimensions.
The surface curvature 
scales with $~n^{1/2}$ in case of quasi 2D and with $~n^{1/3}$ in case of 3D, where $n$ is the number of moving particles 
in the simulation. 
Note that owing to the discrete nature of particles and the relatively small system size the curvature of the solid substrate is not smooth but consists 
of steps, with a step height equal to particle diameter, as shown in Figure \ref{fig2}. Because of these finite steps, the three-phase contact line will be in contact
with different crystallographic axes in each simulation. Different crystallographic axes exhibit different surface energies which may lead to a slight change in the
contact angle\cite{penn1998,steinhart2003}. We have ignored this effect as we have averaged the contact angle for different 
equilibrium profiles which means that the contact angle calculated from our simulations
is average over different crystallographic axes. We have also ignored the effect of surface reconstruction as solid particles remain fixed in an fcc lattice 
during the whole simulation. This occurs either with less stable metal surfaces, semiconductor surfaces at very high temperatures or with polymer surfaces with polar groups
\cite{tretinnikov1994,wang1991,binnig1983}. 
We are not dealing with polar polymer groups or high temperatures so it is justified to ignore this effect.

\begin{figure}

\includegraphics[width=150mm]{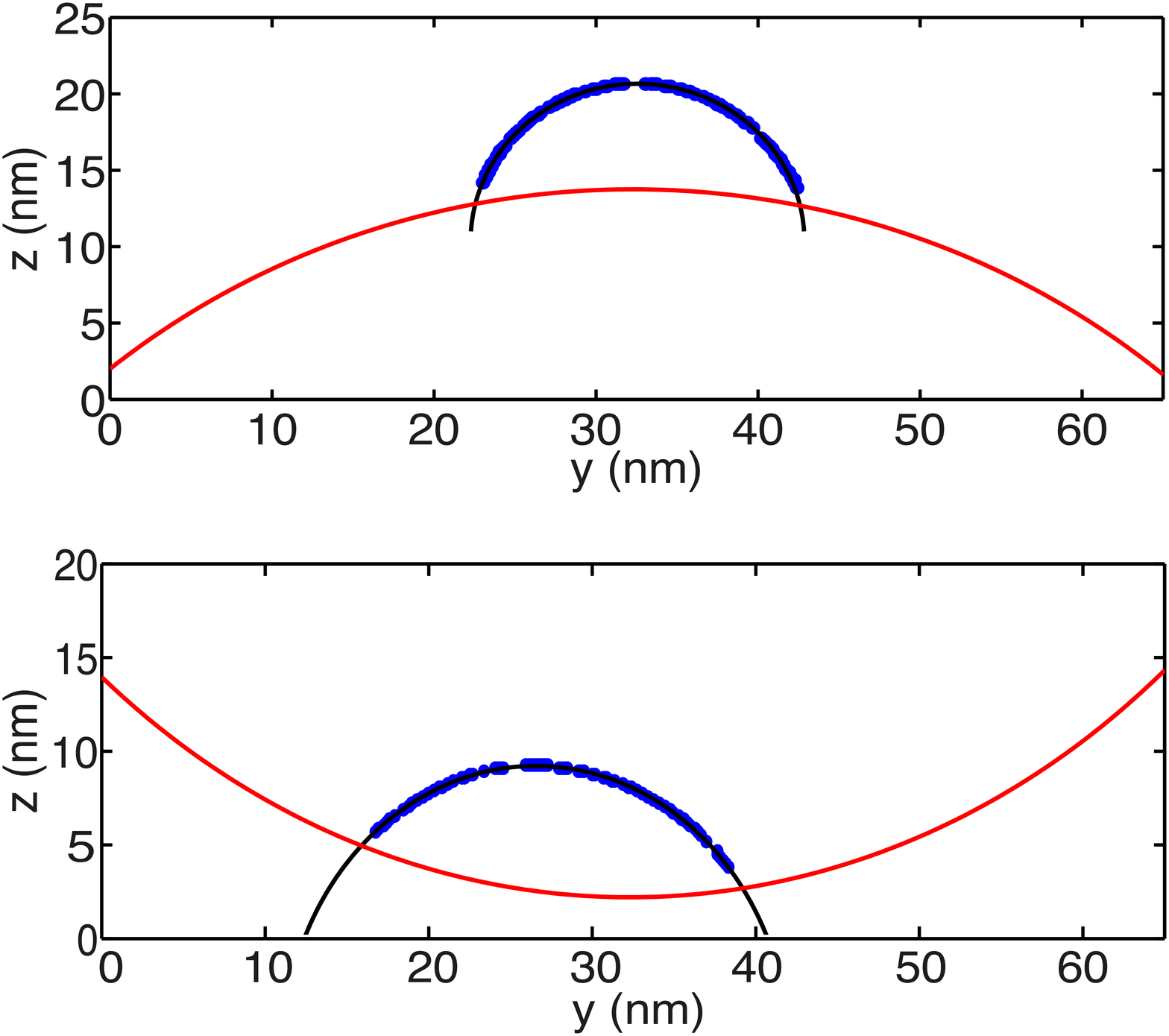}
\caption{Example of a circular cap fit (black line) to the iso-density contour of 0.5 (points) for  a drop on a convex and a concave surface of
constant curvature. The drops can freely shift on the surface due to the statistical fluctuations. In particular, in the concave case the left-shift 
as compared to the initial situation (Figure \ref{fig2}, bottom left) is apparent.}
\label{circlefit}
\end{figure}

Initially, the liquid particles are set in a fcc configuration close to the solid substrate, and free to move from then on at the prescribed temperature (see Figure \ref{fig2}). 
After equilibrium has been reached, i.e. after around $5\times10^{7}$ time steps, 
the density field is calculated by averaging over typically $1,000,000$ time steps (which corresponds to roughly $~2$ nanoseconds)
taking into account the fluctuation of the center of mass of the droplet.
The radius of curvature of the droplet is then obtained by fitting a sphere (circle in 2D) to the iso-density contour of $0.5$ of the normalised density field,
$\rho^{*}(r)$, defined as $\rho^{*}(r) = \frac{\rho(r)-\rho_{V}}{\rho_{L}-\rho_{V}}$, where $\rho_{V}$ and $\rho_{L}$ is the bulk vapour and liquid density,
respectively. 
Since the liquid very near to the solid 
substrate is subject to layering, we have excluded
the density field in the range of $2\sigma$ from the substrate for the circular cap fitting.
From the intersection of the circular fit with the substrate, the contact angle and volume of the drop are evaluated (see figure \ref{circlefit}).
Note that we have splitted the time interval over which we measured into 10 subsets, and calculated the average of 
each subset in order to  evaluate a standard deviation, from which the error bars in the results of Figures \ref{angle2d}, \ref{vdrop2d} and \ref{rdrop2d} were obtained. 

\section{Results and discussion}

\subsection{Wettability of cylindrical nanodrops on curved surfaces}
\begin{figure}
  \includegraphics[width=150mm]{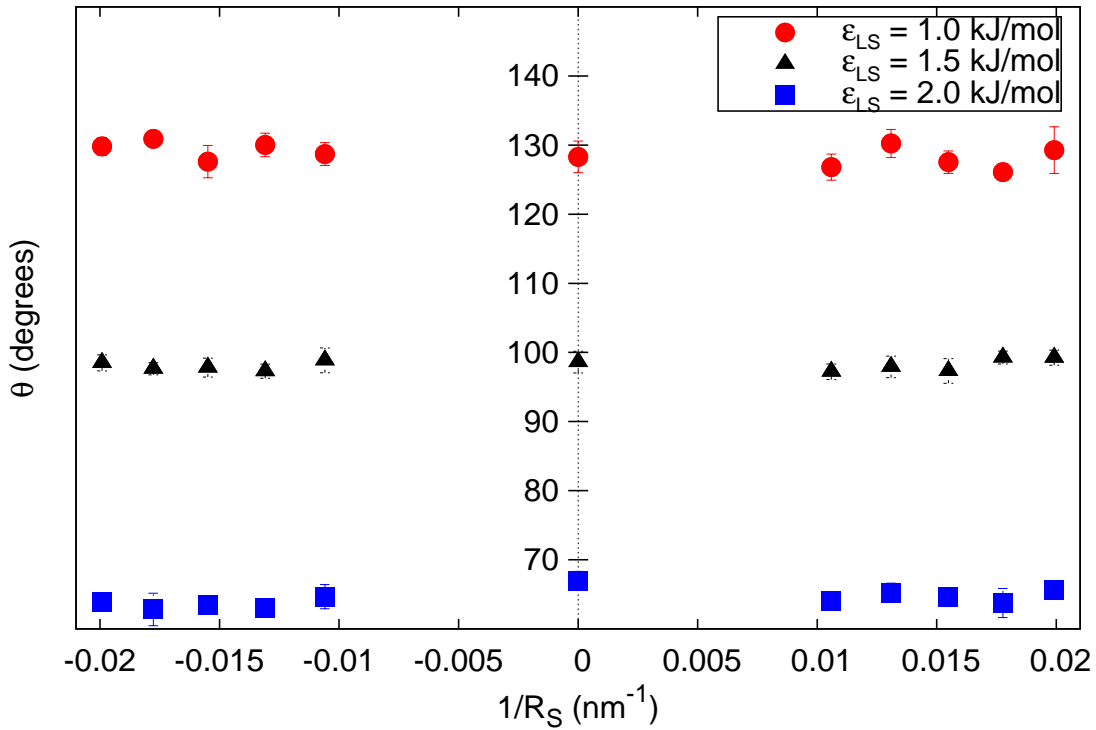}
  \caption{Contact angle $\theta$ of the nanodrop on a solid substrate as a function of surface curvature $1/\rm{R_{S}}$.}
  \label{angle2d}
\end{figure}

\begin{figure}
  \includegraphics[width=150mm]{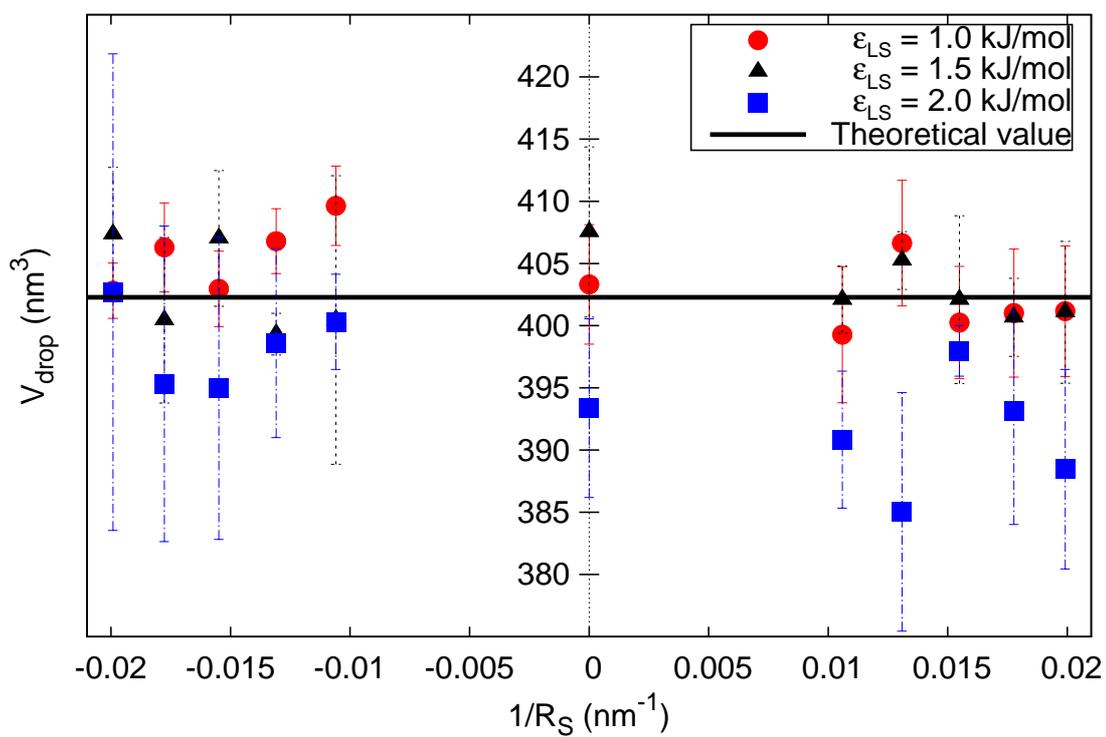}
  \caption{Volume of the nanodrop on a solid substrate as a function of surface curvature $1/\rm{R_{S}}$.}
  \label{vdrop2d}
\end{figure}

\begin{figure}
  \includegraphics[width=150mm]{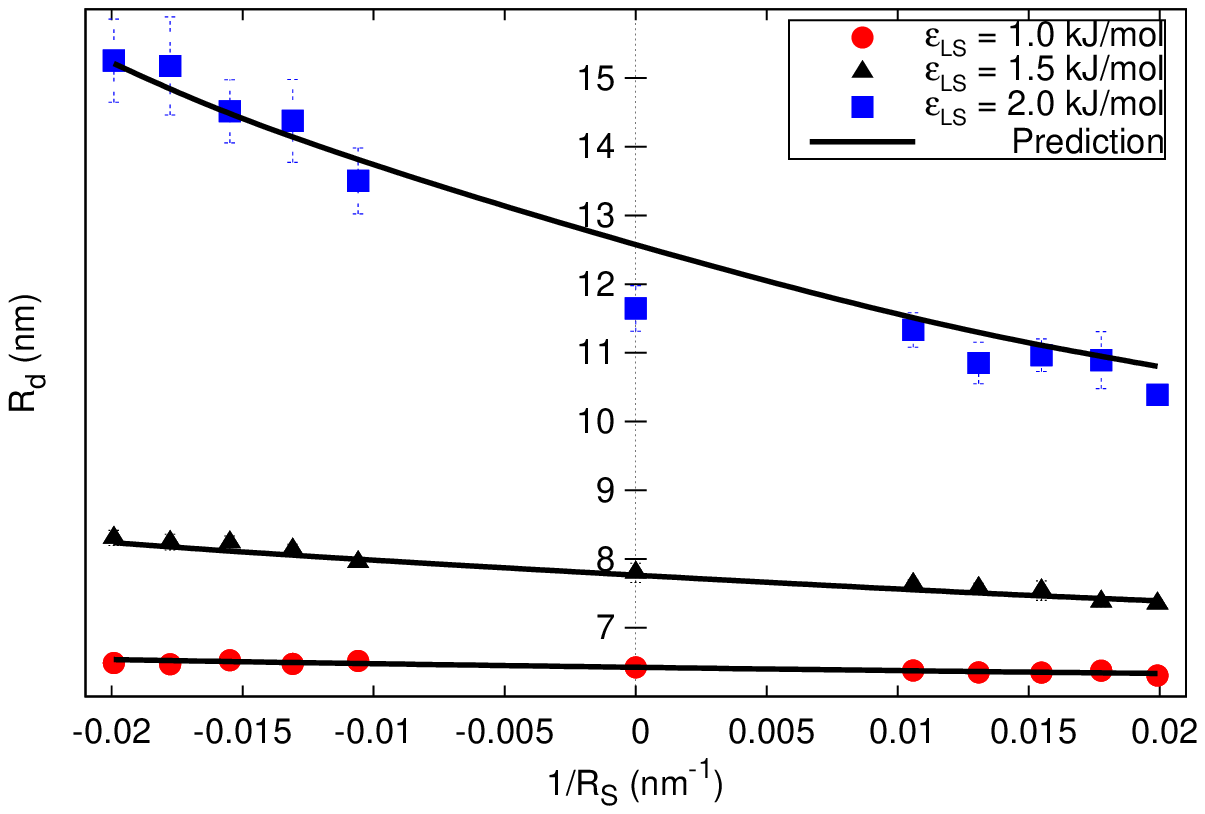}
  \caption{Radius of curvature $R_{d}$ of the nanodrop on a solid substrate as function of surface curvature $1/\rm{R_{S}}$.}
  \label{rdrop2d}
\end{figure}

Figure \ref{angle2d} shows the Young contact angle as a function of the surface curvature, for three different values of the liquid-solid interaction strength.
It can be clearly seen that the contact angle is unaffected by the surface curvature, which is consistent with the 
experimental observations by \citeauthor{extrand2008}\cite{extrand2008} for microscopic drops.
Note that \citeauthor{wolansky1998}\cite{wolansky1998} also showed theoretically that the contact angle is independent of the shape of the surface,
unless line tension is considered. This means that curvature effects are only prominent in nanoscale systems where the line tension is 
appreciable as shown in the next section.
Figure \ref{vdrop2d} shows that also the drop volume does not change with the surface curvature, which is expected since the volume is set by the 
condition of liquid-vapour equilibrium, which to first-order is not affected by the surface curvature. In fact, the  straight line in Figure \ref{vdrop2d} 
is the volume of the liquid drop evaluated from the bulk vapour-liquid equilibrium calculated from a highly accurate
equation of state of the LJ fluid for the given temperature and overall number density\cite{johnson1993}.
The slight difference with the volume as found in the simulations could be attributed to the effect of the liquid-vapour surface and the
solid substrate, which are not accounted for in bulk phase equilibrium. 
Figure \ref{rdrop2d} shows the variation of the radius of drop curvature $R_{d}$ with surface curvature $1/\rm{R_{S}}$. The simulation results 
are found to be in very good agreement with the prediction for $R_{d}$ that follows from straightforward geometric relations 
for the drop volume as function of the contact angle and surface curvature.
In this, we take the volume of the drop as calculated from the equation of state, and the contact angle for a planar surface as reference values.

\subsection{Line tension of nanodrops on curved surfaces}

\begin{figure}
  \includegraphics[width=150mm]{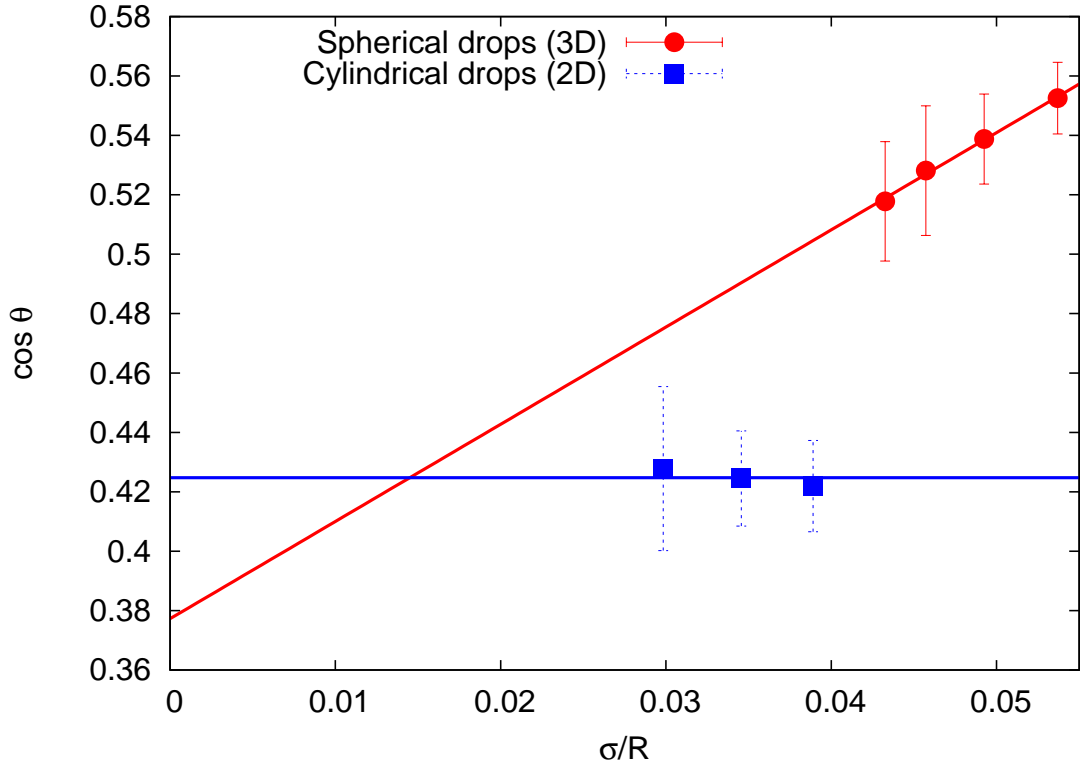}
  \caption{Results from the simulation for the contact angle on a concave surface with $V_{drop}^{\frac{1}{3}}/\rm{R_{S}}$ equal to $-0.2$
  as a function of the curvature of the base circles, $\sigma/R$. The circles are the 
  results for spherical drops, while the squares are for cylindrical drops. Straight lines are the linear fit through 
  the data points. The slope of the straight line for spherical drops gives the line tension length $\ell=-\tau/\gamma_{LV}$.}
  \label{slopetl}
\end{figure}

The line tension of LJ nanodrops has been calculated by fitting the modified
Young equation, eq \ref{myl}, to the simulation data. To this end, different sized LJ drops were simulated in 3D on the curved surface, and the cosine of
the contact angle is plotted against the inverse of the radius of the curvature of the three phase contact line, $1/R$, as shown in Figure \ref{slopetl}.
The slope of the straight line fitted through the data points then gives the line tension length $\ell = -\tau/\gamma_{LV}$. To double-check, 
in Figure \ref{slopetl} we also show the result for $\cos\theta$ for the quasi-2D system (cylindrical drop), which as expected
is not changing with the curvature of drop because the contact line is free of any curvature and the line tension does not have any effect on it.
According to eq \ref{myl}, the two lines in Figure \ref{slopetl} should intersect at zero surface curvature. However, we find a small offset, 
which can most likely be contributed to the fact that the exact position of the liquid-vapour interface is not well defined. 
That is, there is a smooth transition between the two distinct phases and many logical definitions are available to calculate the position of the interface.
Hence quantities which are derived from the interface location, such as contact angle, volume, radius of curvature of drop, etc., 
will slightly vary depending on the definition,\cite{schimmele2007}, which may result in a slight offset from eq \ref{myl}.
Note also that for these typical values of the contact angle, only a slight error in  $\theta$ of say $1\%$ leads  to errors of $3\%$ in $\cos\theta$,
which leads even to larger errors in the extrapolated value of the line fit. 

\begin{figure}
  \includegraphics[width=150mm]{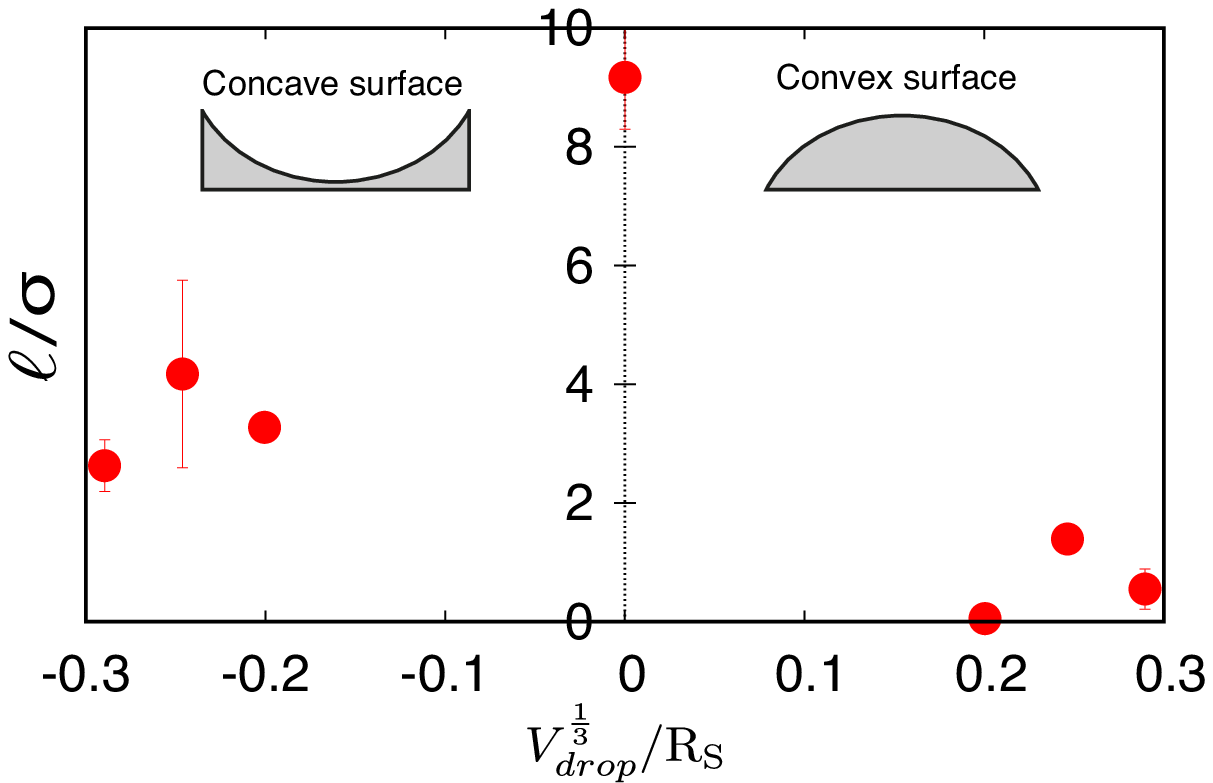}
  \caption{Variation of line tension length $\ell$ as function of $V_{drop}^{\frac{1}{3}}/\rm{R_{S}}$.
  Note that smaller moduli of $V_{drop}^{\frac{1}{3}}/\rm{R_{S}}$ are not possible due to the graininess of the surface particles.}
  \label{tensionl}
\end{figure}

\begin{figure}
  \includegraphics[width=150mm]{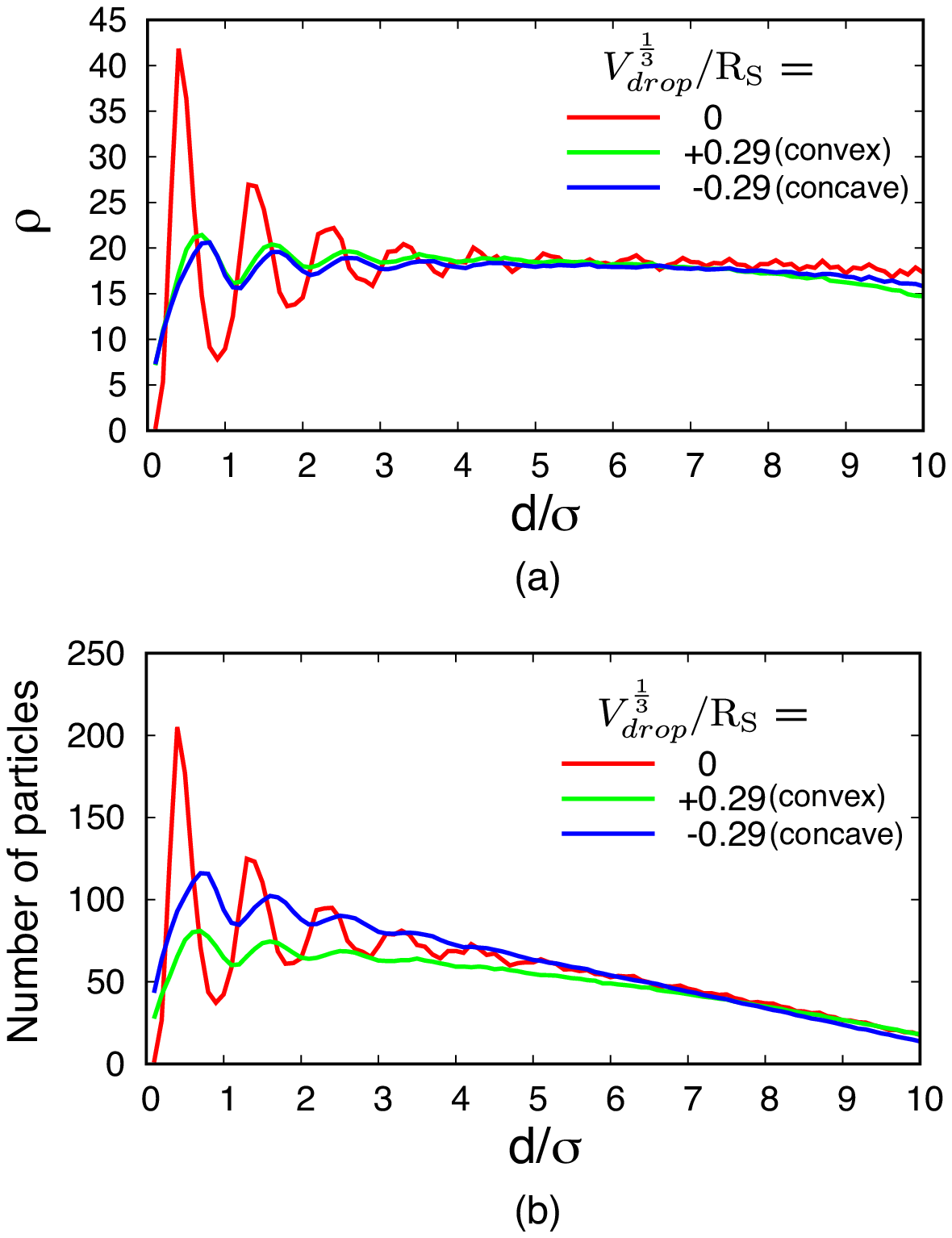}
  \caption{(a) Density and (b) absolute number of particles in the drop as function of distance from the surface. Density of particles is calculated by dividing
  the absolute number of particles by the volume of each layer.}
  \label{den_num_layer}
\end{figure}

\begin{figure}
  \includegraphics[width=150mm]{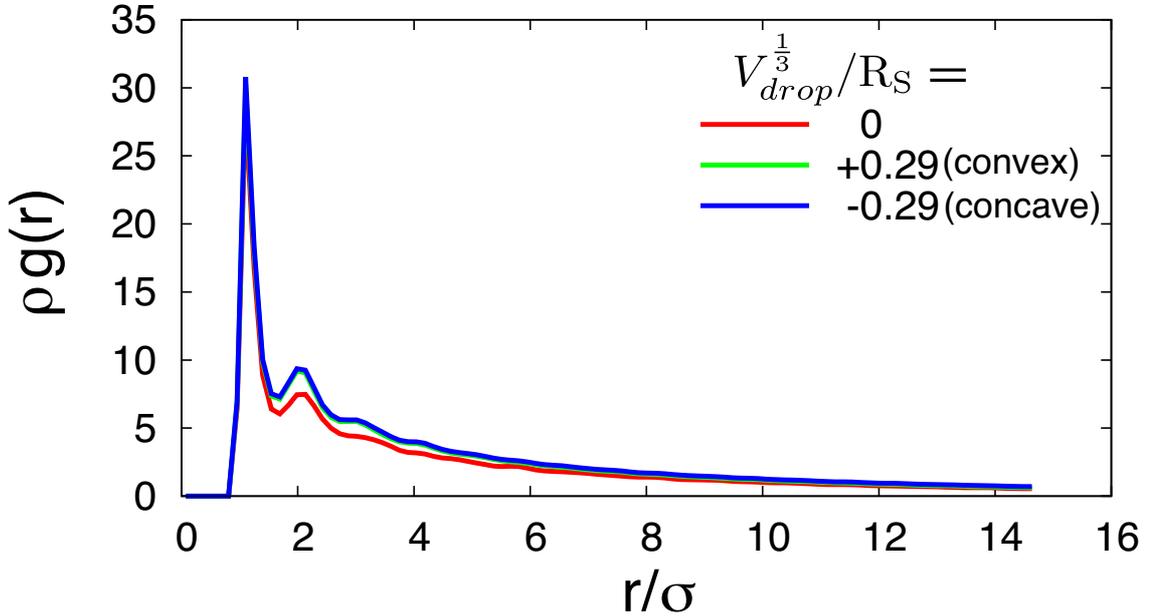}
  \caption{Radial distribution function of particles with in two layers from the surface.}
  \label{rdf}
\end{figure}

\begin{figure}
  \includegraphics[width=150mm]{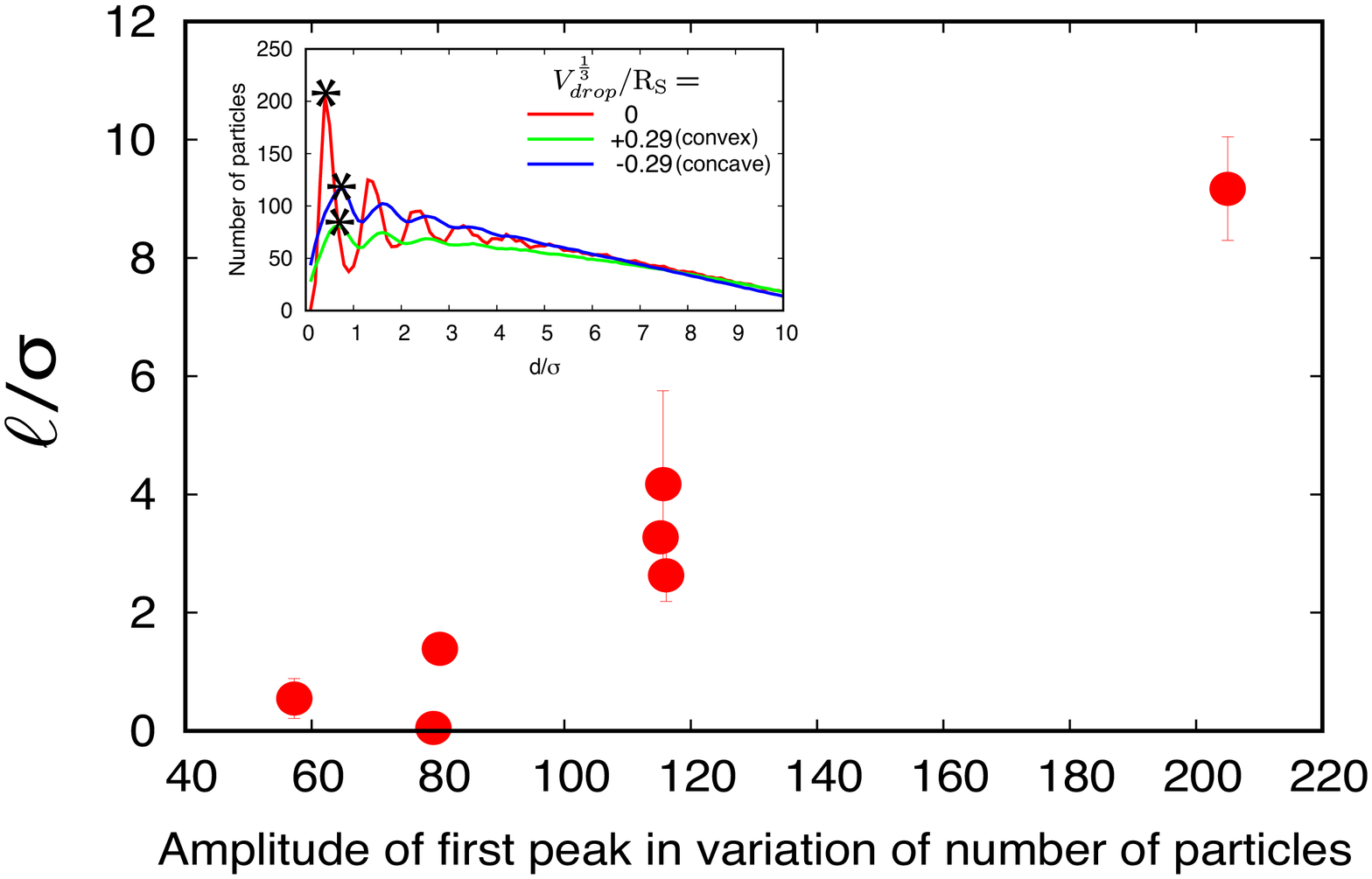}
  \caption{Amplitude of the first peak in oscillations of variation of number of particles as function of the distance from the surface is plotted against the magnitude of the 
  line tension length for various surface curvatures. 
  The asterisk in the figure shown in inset shows the peaks which are plotted in the graph.}
  \label{amplitude}
\end{figure}

Figure \ref{tensionl} shows the magnitude of the line tension length as a function of the surface curvature
normalised by $V_{drop}^{\frac{1}{3}}$, where $V_{drop}$ is the volume of the drop as calculated from the equation of state
for Lennard-Jones fluid\cite{johnson1993}.
We find a clear maximum in case of planar surface which is 
compatible with the theoretical predictions\cite{marmur2002,wolansky1998}, however surprisingly we 
find that $\ell$ is much higher in case of concave surfaces (negative curvature)
as compared to convex surfaces (positive curvature). From Figure \ref{tensionl}, we can infer that the magnitude of the surface curvature does not have a very
strong effect on the line tension length but it is strongly dependent on the sign of the surface curvature. We investigated this finding by analysing the 
arrangement of particles very close to the surface. In Figure \ref{den_num_layer}, we have plotted the variation of density of particles 
and the absolute number of particles in a drop as a function of distance from the surface.
Both quantities are evaluated from the time-averaged number of particles for concentric spherical shells or layers of thickness $0.1\sigma$.
The variation in the absolute number of particles is determined by averaging the number of particles in each layer over time.
Density in each layer is then calculated by dividing the absolute number of particles in each layer by the volume of that layer.
It can be seen that the amplitude of the oscillations in the density is much larger in the case of the planar surface as compared to the curved ones, yet the
difference in density oscillations between positive and negative surface curvature is small. However, the variation in the number of particles in the drop
as a function of distance from the surface  clearly shows the difference between the three types of surfaces.
The amplitude of the first peak is maximal in case of a flat surface, then followed by the peak of negative and positive surface curvature, respectively. 
This trend is directly correlated with the magnitude of the line tension length, which suggest that more particles in layers close to the surface imply a larger
line tension. 
Note that the absolute number of particles and the density of particles as a function of distance are related to each other by the volume of each layer. 
Since the density of particles is almost 
the same for both curvatures, this implies that volume of layers very close to the surface is larger in case of a concave surface.
We have also analysed the structure of the particles very near to the surface by evaluating radial distribution functions or pair correlation functions
for particles in the first two layers as shown in Figure \ref{rdf}. The magnitude of the second peak in the 
radial distribution function is marginally larger in case of curved surfaces as compared 
to the plane surface, but there is no difference between positive and negative surface curvatures which indicates that it is the absolute number of particles in the layers
near to the surface which is responsible for the decrease in magnitude of the line tension and not the relative arrangement of the particles.
From Figure \ref{amplitude}, we can see a direct proportionality between the line tension length and the amplitude of the first oscillation of variation of number of particles
with the distance from the surface.

The actual quantity of interest, the line tension, can be calculated by multiplying the line tension length by the liquid-vapour surface tension $\gamma_{LV}$.
\citeauthor{kirkwood1949}\cite{kirkwood1949} showed that the latter can be calculated by integrating
the difference between normal and tangential components of the pressure tensor across the interface,
\begin{equation}
\gamma_{LV} = \frac{1}{2}\int_0^{L_{z}} [p_{n}(z)-p_{t}(z)]\mathrm{d}z.\label{buff}
\end{equation}
We have used this procedure to calculate $\gamma_{LV}$ from independent molecular dynamics simulation of liquid molecules in equilibrium with its own vapour and
for a planar interface. 
Note that the factor $1/2$ in eq \ref{buff} is the correction for the extra interface that is present due to periodic boundary conditions. We again emphasise that 
we have not considered surface tension as a function of interface curvature.
The radius of curvature of nanodrop of 7 nm is required to change the surface tension by 5\% for the Lennard-Jones particles that we are simulating
\cite{tolman1949,nijmeijer1992}.
The radius of curvature of drops in our simulations
is in the range of $~10$ nm.
So the assumption of constant surface tension is fairly acceptable.
For our Lennard-Jones parameters, i.e. $\epsilon_{LL} = 3.0$ kJ/mol, 
this procedure gave a value of $\gamma_{LV}=4.1893\times10^{-3}$ N/m. Using this value for the surface tension, the order of magnitude of the line tension 
is coming in the vicinity of $10^{-12}$ N which is very close to the theoretical prediction and many experimental findings
\cite{marmur1997,schimmele2007,getta1998,dobbs1993,amirfazli2004}. The maximum value of the line tension is 
around $13\times10^{-12}$ N for planar surface and the minimum value is $0.07\times10^{-12}$ N, which is almost zero, in case of convex surface.

\section{Conclusions}
Molecular dynamics simulations were performed for liquid drops on curved surfaces. The Young contact angle was found to be constant 
with surface curvature which is consistent with previous experimental and theoretical predictions. The volume of the drop is also found to be
independent of the curvature of the surface, and the radius of curvature of the drop is well 
predicted by simple geometric relations by keeping contact angle and volume constant.
The magnitude of the line tension is calculated by MD simulation and found to be comparable with theoretical values. The line tension
strongly depends on the sign of the surface curvature: its value is much larger in case of negative curvature (concave surfaces) as compared to
positive curvature (convex surface), while it reaches a maximum for zero curvature. This trend is found to be correlated with the number of particles
in the initial layers of the drop.
The relative arrangement of particles near the surface is found to be the same which means that the number of particles near the 
three phase contact line is the only significant factor which is responsible for altering the magnitude of the line tension on curved surfaces.

Authors thank Joost Weijs for useful discussions and help in MD simulations using GROMACS
and SURFsara and NWO for providing computational facilities for the simulations and FOM
for financial support.

\bibliographystyle{unsrtnat}
\bibliography{references}

\end{document}